\documentstyle[preprint,revtex]{aps}
\draft
\begin{document}
\begin{title}
{Scaling behavior of the four-point renormalized
coupling constant in the two dimensional O(2) and O(3)
non-linear sigma models}
\end{title}
\author{Jae-Kwon Kim}
\begin{instit}
{Department of Physics, University of California,
Los Angeles, CA 90024, and \\
Department of Physics, University of Arizona, Tucson, AZ 85721}
\end{instit}

\begin{abstract}
We report thermodynamic values of four-point renormalized
coupling constant calculated by Monte Carlo simulations
in the continuum limits of
the lattice versions of the two-dimensional O(2) and O(3)
non-linear sigma models. In each case the critical index of
the coupling constant vanishes, which leads to hyperscaling (
non-triviality).
\end{abstract}

A  general problem of (Euclidean) quantum field theory (QFT)
is whether the n-point ($n>2$) correlation functions, which are
constructed so as to satisfy Osterwalder-Schrader axioms\cite{OST},
are those of the Gaussian model in the continuum limit.
If so this QFT is non-interacting,
and is referred to be a {\it trivial} theory\cite{SOK}.
It is rigorously proven that one cannot
construct an interacting continuum limit of the $\lambda \phi^{4}$
theory in the symmetric phase in dimensions larger than 4
($D > 4$)\cite{AIZ}, and is widely conjectured that this is
the case even in four dimensions\cite{HARA}.
A trivial theory, which might be interacting only when a finite
cutoff is imposed in the theory, cannot be a candidate of a  genuine QFT,
so it is very important to find out
whether a theory is trivial or not.
Without any proof, it is generally believed that a QFT
which is not asymptotic free in 4D would be trivial. An important
example of such a theory is the 4D non-compact QED, and
the issue of triviality in this theory remains unresolved with
variant conclusions of studies.

As far as the issue of the triviality is concerned,
studying the scaling behavior of the four-point
renormalized coupling constant ($g^{(4)}$) is primarily important.
A rigorous theorem\cite{NEW} states that vanishing of $g^{(4)}$ is sufficient
for triviality in the Ising like ferromagnetic systems,
and it is conjectured that the same  holds for other type of
ferromagnetic lattice models\cite{SOK}.
While much (numerical) studies on the scaling behavior of $g_{R}^{(4)}$
have been done for the $\lambda \phi^{4}$ theories
in various dimensions\cite{WEIN}\cite{KIMD} and
for the Yukawa type interaction\cite{HAS},
little has been studied for the
2D O(N) models for $N \ge 2$\cite{LUS}, which share many similarities
with 4D lattice gauge theories.
In particular, the 2D O(2) non-linear sigma model is not
asymptotic free in the context of perturbation theory, while
O($N\ge 3$) non-linear sigma models are asymptotic free\cite{POL}.

In this article, we investigate the scaling behaviors of $g^{(4)}$
for the 2D O(N) (N=2,3) non-linear sigma models on
a square lattice of linear size L  with periodic boundary condition,
based on Monte Carlo simulations.
The model is defined by the action
\begin{equation}
A=-\beta \sum_{<i,j>} \sigma_{i} \cdot \sigma_{j},
\end{equation}
where $\beta$ is the inverse temperature ($\beta=1/T$), and
$\sigma_{i}$ is an N-dimensional unit vector
at lattice site $i$.
We report compelling  numerical evidences that the continuum limits
of both models are not trivial, irrespective
of the property of perturbative asymptotic freedom.

The four- point renormalized coupling constant ($g^{(4)}$)
which is basically the  connected
four-point function defined at zero-momentum, and which
is constructed
so as to be independent of rescaling of field strength,
can be written in a system with translation symmetry as
\begin{equation}
g^{(4)} = \lim_{L \to \infty} g_{L}^{(4)}\equiv
\lim_{L \to \infty} U_{L}~~ (\xi_{L}/L)^{D}, \label{eq:def}
\end{equation}
Here, $g_{L}^{(4)}$ and $\xi_{L}$ are respectively
the renormalized coupling constant and correlation length\cite{DEF}
defined  on a finite lattice of linear size L,
and $U_{L}$ is a modified (N component)
Binder's cumulant ratio  defined as
$U_{L}\equiv [(1+2/N)<S^{2}>^{2}-)<S^{4}>]/<S^{2}>^{2}$,
with $S^{2}\equiv |\sum_{i} \sigma_{i}|^{2}$\cite{BER}.

For a system displaying power-law type scaling behavior,
$g^{(4)}$ scales as
$g^{(4)}(t) \sim t^{-2\Delta+\gamma+D\nu}$, where the
notations are standard ($t$, for example, is the deviation of
the dimensionless temperature from the coupling defined as
$t\equiv (\beta_{c}-\beta)/\beta_{c}$).
The inequality, $2\Delta \le \gamma+D\nu$ holds\cite{BAK}
in general for ferromagnetic systems with power-law singularities.
For the $\lambda \phi^{4}$ models, the inequality
has been rigorously proven for $D >4$ so that
$\lim_{t \to 0} g^{(4)} \to 0$, whereas the equality
holds for $ D \le 4$ (hyperscaling)\cite{AIZ}.
In 4D, it is rigorously proven under some mild
assumptions\cite{HARA} that the scaling behavior of $g^{(4)}(t)$
has such a multiplicative  logarithmic correction  that
$\lim_{t \to 0} g^{(4)}(t) \to 0$.

For the 2D O(N) (N=2,3) models,
we have calculated $g^{(4)}$
by employing Wolff's single cluster algorithm \cite{WOL}
on square lattice with periodic boundary condition.
For each model, our continuum limit is achieved by adjusting
the value of $\beta$ (or, $T$) so that the correlation length in unit
of lattice spacing starts to be diverging; at the same time, the lattice
spacing scales with $\beta$ so that the correlation length
multiplied by the lattice spacing remains a constant.
In general the auto-correlation time for $g^{(4)}$ is much
larger than that of $\chi$ or $\xi$, so much more computational
efforts are required for the precise determination of $g^{(4)}$.
For a given $\beta$ and L, 20-80 different {\it bins} were
obtained for our calculations, with each bin being
composed of 10 000 measurements
each of which was separated by 8-15 consecutive one cluster
updating. The statistical errors were estimated by jack-knife method.

In order to monitor the effect of finite size
in the measurements of $g^{(4)}$, for each model
at an arbitrary temperature in the scaling region, we
measured $g_{L}^{(4)}$ by varying L by 10  from
L= 10 until $g_{L}^{(4)}$ did not vary with further increasing of L.
For each model, we conclude that $g_{L}^{(4)}$ (as well as
$\chi$ and $\xi$) converges to
its thermodynamic value (within the statistical error of
very precise Monte Carlo data) on the condition that
$L /\xi_{\infty} \ge 7$ (Figure(1)).
According to the theory of finite size scaling\cite{KIML},
the following relation holds for the renormalized four-point coupling:
\begin{equation}
g_{L}^{(4)}(\beta)= g^{(4)}(\beta) f_{g}(s),~~s\equiv L/\xi_{\infty}(\beta)
\label{eq:fun},
\end{equation}
with  $f_{g}$ representing
a scaling function characterized by $g^{(4)}$.
$f_{g}(s)$ has no explicit
temperature dependence so that the {\it thermodynamic} condition
holds for any temperature.

We thus chose L such that $ L /\xi_{\infty} \simeq 7$ for the
direct measurement of the thermodynamic $g^{(4)}$
(T=1.19, 1.10, 1.04, and
1.02 for the O(2) model; $\beta=1.5, 1.6$, and 1.7 for the O(3) model).
For the temperatures where the corresponding $\xi$ becomes
very large, we fix $L/\xi_{\infty}$ to be a value much
smaller than 7 so that, according to Eq.(\ref{eq:fun}),
$g_{L}^{(4)}$ thus obtained is exactly
proportional to its corresponding thermodynamic value.
When the corresponding $\xi_{\infty}(\beta)$ is known
at a certain $\beta$,
this procedure enables one to obtain accurate thermodynamic values
of other physical quantities without using sufficiently large L
required in direct measurements\cite{KIML}.
Let us illustrate this for the 2D O(2) model, where very accurate
values of $\xi_{\infty}$ are already available up to $T=0.98$\cite{GUP}.
At $T=1.02$ ($\xi_{\infty}=26.20(20)$),
we  obtained $g^{(4)}=8.86(13)$ using L= 184
and $g_{55}^{(4)}=4.79(2)$, so $f_{g}(s)$ is calculated to
be 0.541(10) for $s= L/\xi_{\infty}=2.10(2)$.
At $T=0.98$ ($\xi \simeq 70.5(7)$),
choosing L=148 makes the value of $s$ the same as
at $T=1.02$ within the statistical errors,
so  with our Monte Carlo data, $g_{148}^{(4)}=4.81(2)$,
we extract $g^{(4)}=8.89(20)$ for $T=0.98$.
For  the O(3)  model, we fixed the value of $s=1.165(21)$ by choosing
L=40, 75, and 142 for $\beta=1.7, 1.8,$ and 1.9 respectively.
Our final results are summerized in Table (1).

For such broad ranges
of correlation length ($5.01(2) \le \xi \le 70.5(9)$ for the
O(2); $11.1(1) \le \xi \le 121.9(7)$ for the O(3)) we observe that
$g^{(4)}(t) \sim t^{0}$. These behaviors are qualitatively the
same as those in the 2D and 3D Ising models\cite{KIMD}.
We thus conclude that the continuum limits are non-trivial
for both models. In particular, the perturbative property of the
non-asymptotic freedom in the 2D O(2) model and that of
the asymptotic freedom in the 2D O(3) model do not affect
the non-trivialities in these models.
Especially, because of the asymptotic
freedom in the 2D O(3) model,
the value of the renormalized coupling constant defined at any other value
of the momentum must be smaller than our value, $g^{(4)}=6.6(1)$.
Another remarkable conclusion of the current study is
the existence of hyperscaling in these models.
Hyperscaling relations cannot be defined in our models
due to the exponential singularities in the critical
behaviors of $\xi$ and $\chi$; nevertheless,
the fundamental claim of the hyperscaling in the sense that the
thermodynamic correlation length is the only relevant scale in the
scaling region,  still holds in two dimensions
irrespective  of the type of critical behavior.

The author would like to thank Chuck Buchanan for hospitality;
Ghi-Ryang Shin and Jaeshin Lee for their support in computing.
The Monte Carlo simulations were carried out on Convex C240
at the University of Arizona.

\begin{table}
\caption{$g^{(4)}$ for the 2D O(2) and O(3) non-linear sigma models.
For the O(3) model, we fixed $s=L/\xi_{\infty}=1.165$ by choosing
L=40, 75, and 142 for $\beta=1.7, 1.8,$ and 1.9 respectively.
The corresponding values of $g_{L}^{(4)}$ are 2.56(1), 2.56(1), and
2.59(2) respectively, showing the constancy of $g^{(4)}$ in this
range of $\beta$ as well. $f_{g}(s)$ is calculated to be 0.394(5).
$\xi_{\infty}(\beta)$ =64.6(5) and 121.7(8) for $\beta$=1.8 and 1.9
\cite{FOX}.}
\begin{tabular}{ccccccc}
O(2)  &T         &1.19     &1.10     &1.04   &1.02 &0.98  \\
      &$g^{(4)}$ &8.70(16) &8.80(12) &8.71(10) &8.86(13) &8.89(20) \\
	\hline
O(3)  &$\beta$   &1.5	   &1.6	 	&1.7	&1.8	&1.9 	\\
	&$g^{(4)}$ &6.58(13) &6.70(15) &6.50(6) &6.50(11) &6.57(13)\\
\end{tabular}
\end{table}

\vspace{2cm}
{\bf Figure Caption} \\
Figure(1): $g_{L}^{(4)}/g^{(4)}$ as a function of $L/\xi_{\infty}$
	   at $T=1.10$ ($\xi_{\infty}=9.32(2)$) for the 2D O(2) model,
		and at $\beta=1.5$ ($\xi_{\infty}=11.05(2)$) for the 2D
		O(3) model. The range of L is [10,80] for the O(2),
	        and is [10,100] for the O(3), varied by 10.

\end{document}